\documentclass[aps,pre,twocolumn,groupedaddress,showpacs,floatfix]{revtex4}
\usepackage{amsmath}
\usepackage{graphicx}
\usepackage{dcolumn}
\usepackage{bm}
\usepackage{epsfig}
\input epsf
\epsfclipon
\begin{document}
\title{Publish or perish: analysis of scientific productivity using
maximum entropy principle and fluctuation-dissipation theorem}
\author{Piotr Fronczak, Agata Fronczak and Janusz A. Ho\l yst}
\affiliation{Faculty of Physics and Center of Excellence for
Complex Systems Research, Warsaw University of Technology,
Koszykowa 75, PL-00-662 Warsaw, Poland}
\date{\today}

\begin{abstract}

Using data retrieved from the INSPEC database we have
quantitatively discussed a few syndromes of the publish-or-perish
phenomenon, including continuous growth of rate of scientific
productivity, and continuously decreasing percentage of those
scientists who stay in science for a long time. Making use of the
maximum entropy principle and fluctuation-dissipation theorem, we
have shown that the observed fat-tailed distributions of the total
number of papers $x$ authored by scientists may result from the
density of states function $g(x;\tau)$ underlying scientific
community. Although different generations of scientists are
characterized by different productivity patterns, the function
$g(x;\tau)$ is inherent to researchers of a given seniority
$\tau$, whereas the publish-or-perish phenomenon is caused only by
an external field $\theta$ influencing researchers.
\end{abstract}


\maketitle

\section{Introduction} \label{intro}

Nowadays, {\it ($\dots$) Evaluations of scientists depend on
number of papers, positions in lists of authors, and journals'
impact factors. In Japan, Spain and elsewhere, such assessments
have reached formulaic precision. But bureaucrats are not only
wholly responsible for these changes - we scientists have
enthusiastically colluded. What began as someone else's measure
has become our (own) goal.($\dots$)} \cite{LawNature03}. In fact,
a number of scientists all over the world alter that research is
in crisis. Academics are having to {\it publish-or-perish}.
Scientific articles become a valuable commodity both for authors
and publishers \cite{GadPT04}. The politics of publication does
not only concentrate on publishing as valuable articles as
possible. Of course, since articles in leading journals certifies
one's membership in the scientific elite the impact factor of
journals matters but also the total number of publications is of
great importance since frequent publications allow to sustain
one's career, and are well seen when applying for funds. Authors
have to plan when, how and with whom to publish their results.
Quoting Lawrence \cite{LawNature03}: {\it The ideal time is when a
piece of research is finished and can carry a convincing message,
but in reality it is often submitted at the earliest possible
moment.($\dots$) Findings are sliced as thin as salami and
submitted to different journals to produce more papers.}
Scientists, who are aware of the publish-or-perish phenomenon warn
that research professionalism may be sacrificed in the pursuit of
research grants and fame, or simply for fear of loss of a
position.

In this paper, using data retrieved from the INSPEC database, we
quantitatively analyze two syndromes of the publish-or-perish
phenomenon: continuous growth of rate of scientific productivity
and continuously decreasing percentage of those scientists who
stay in science for a long time.

The paper is organized as follows. In the next section we start
with a simple examination of scientific productivity distributions
for all INSPEC authors together, as it was done by Lotka
\cite{Lotka1926} and Shockley \cite{Shockley1957}. Then, we study
temporal evolution of the scientists. From the whole database we
draw {\it long-life scientists}, i.e. scientists who were doing
research for at least $18$ years. Having such a set of scientists
we divide it into the so-called {\it cohorts} including those who
started to publish in a given year $T$ (i.e.
$T=1975,\;1976,\;\dots,\;1987$). We show that unlike quickly
increasing number of all authors listed in the INSPEC database the
number of long-life scientists, as characterized by year of the
first publication $T$, remains almost constant indicating
decreasing percentage of long-life scientists among all
researchers. We also show that histograms of scientific
productivity $N(x;t,T)$ within $T$-cohorts, measured by the number
of articles $x$, change over time $t$ from almost exponential
(when cohort contains young scientists) to clearly fat-tailed
(when the same cohort includes mature researchers). Additionally,
we observe that the number of articles produced by a
representative of each cohort increases with the square of
seniority $\tau=t-T$ i.e. $\langle x \rangle\sim \tau^2$,
indicating that each cohort possesses fixed {\it acceleration}
parameter $a(T)=\partial^2\langle x\rangle/\partial \tau^2$ which,
on its own turn, quickly increases with $T$. Finally, in Sec. III,
we analyze the observed distributions of scientific productivity
in terms of equilibrium statistical physics. We show that the
fat-tailed histograms $N(x;t,T)$ may result from the inherent
density of states function $g(x;\tau)$ characterizing scientific
community. We also introduce the parameter $\theta(t,T)$, which
has a similar meaning as the inverse temperature $\beta$ in the
canonical ensemble, and describes an external field influencing
scientists. The parameter allow us to quantify the effect of
publish-or-perish phenomenon.

\section{Scientific productivity - fundamental results}

\begin{figure} \epsfxsize=8.5cm \epsfbox{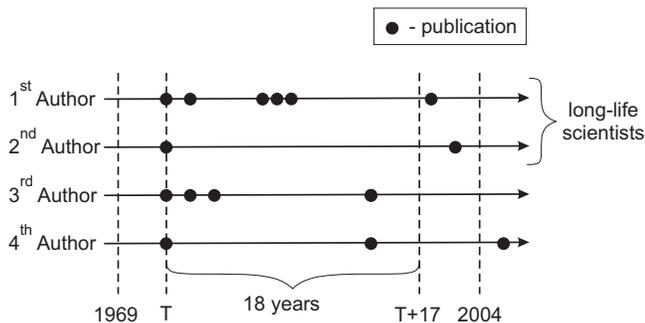}
\caption{The figure explains the procedure used in order to
retrieve long-life scientists. We assume that an author belongs to
the $T-$cohort if the period of time that passed between his/her
first and last publication fulfills the relation $T_f-T\geq 17$,
where $T_f$ is the year of the last publication indexed in our
data set. According to the procedure only the first two authors,
whose publication history is depicted in the figure, are
considered to be long-life $T$-scientists.}\label{fig1}
\end{figure}

In this study we report on scientific productivity of all authors
(over $3$ million) listed in the INSPEC database \cite{inspec} in
the period of $1969-2004$. The database, produced by the
Institution of Electrical Engineers, provides a few million of
records indexing scientific articles published world-wide in
physics, electrical engineering and electronics, computing and
information technology. Although each INSPEC record contains a
number of fields (including publication title, classification
codes etc.) for our purposes we have retrieved only two of them:
authors' names (i.e. names with all initials) and publication
year. Having the data we were able to discover the initial year of
one's scientific activity $T$ (i.e. year of the first publication)
and also the cumulative number of his/her publications in the next
years. Additionally, from the whole data set we have drawn
long-life scientists (i.e. scientists who were productive for at
least $18$ years, see Fig.~\ref{fig1}), and we have divided them
into the so-called $T-$cohorts, with $T$ having the same meaning
as previously.

\begin{figure} \epsfxsize=8.5cm \epsfbox{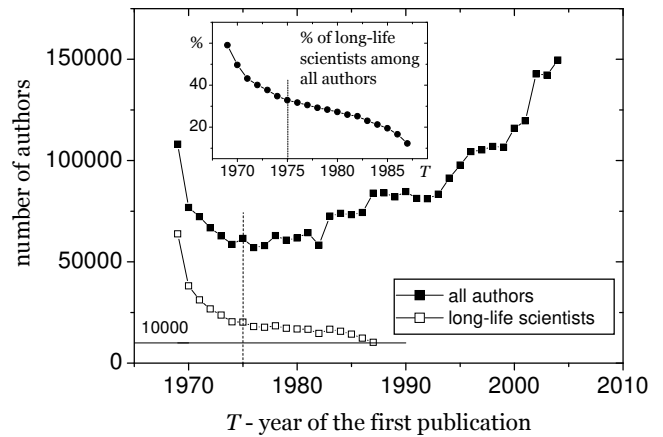}
\caption{Number of all authors listed in the INSPEC database and
the number of long-life scientists versus the year of the first
publication $T$.}\label{fig2}
\end{figure}

A few important findings on evolution of scientific community can
be immediately drawn from the simple comparison of the number of
all $T$-authors and the number of those authors who turned out to
be long-life scientists. However, before we discuss how the
numbers and their ratio depend on $T$, two limitations of our data
should be noted. First, since the INSPEC database does not contain
information about articles published before $1969$, the initial
year of scientific activity $T$ for scientists indexed in the
database in early seventies may be incorrect. That is why, for
further analysis we have restricted ourselves to the period
starting at $T=1975$. Second, due to the the criterion of $18$
years of activity, taken when specifying $T-$cohorts, the number
of cohorts is limited to $13$, respectively for
$T=1975,1976,\dots, 1987$. Keeping in mind the mentioned
constraints one can see (Fig.~\ref{fig2}) that although the number
of all authors listed in the INSPEC database increases every year,
the number of long-life scientists remains almost constant (the
downward trend observed in eighties should not be taken into
account as it may result from finite-size effects due to reduction
of the period between $T+17$ and $2004$; consider the case of the
$2$nd Author in Fig.~\ref{fig1}). The chief conclusion resulting
from the above observations is that the percentage of long-life
scientists among all scientists monotonically decreases in time
(see inset in Fig.~\ref{fig2}).

In the rest of the section we will concentrate on the fundamental
features of distributions describing scientific productivity of
authors indexed in INSPEC. As a matter of fact, scientific
productivity, measured by the number of papers authored, has a
long history of study in socio- and bibliometrics, with the
articles by Lotka \cite{Lotka1926} and Shockley
\cite{Shockley1957} being famous early examples. Both of these
authors found that the number of papers produced by scientists has
a {\it fat-tailed} distribution, exhibiting both a large number of
authors who contributed only a few articles, and a small number of
authors who made a very large number of contributions. Being more
precise, Lotka (1926) studied a sample of $6891$ authors listed in
{\it Chemical Abstracts} during the period of $1907-1916$ finding
that the number of authors making $x$ publications was described
by a power law
\begin{equation}\label{powerlaw}
N(x)\sim x^{-\gamma}
\end{equation}
with $\gamma\simeq 2$, whereas Shockley (1957) investigated
scientific productivity of $88$ research staff members at the
Brookhaven National Laboratory in the USA finding log-normal
distribution
\begin{equation}\label{lognormal}
N(x)\sim\frac{1}{s\sqrt{2\pi}x}e^{-(\ln x-m)^2/(2s^2)}.
\end{equation}

\begin{figure*}
\centerline{\epsfig{file=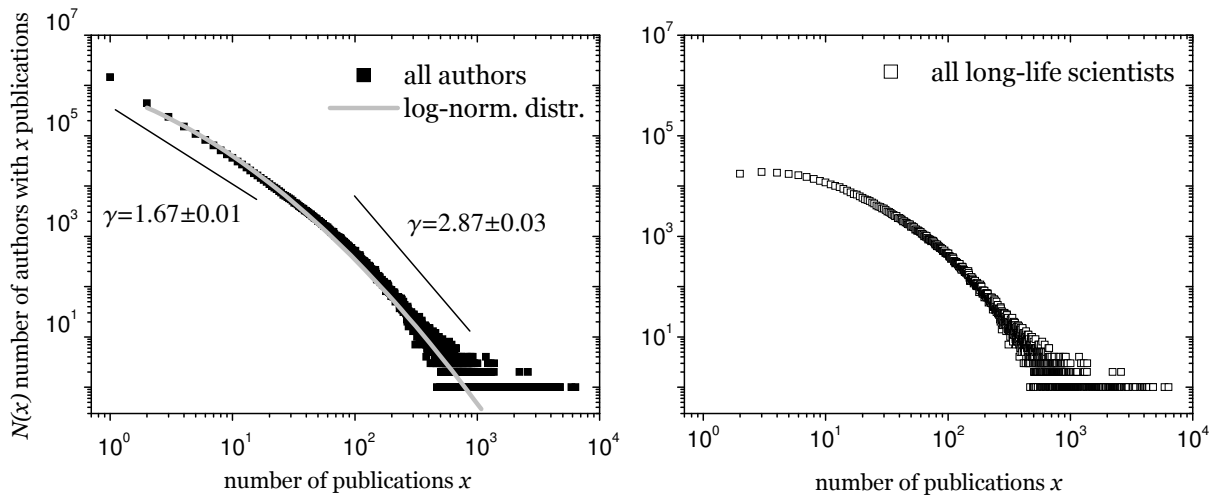,angle=0,width=16cm}}
\caption{Histograms of the number of papers written by: all
authors in INSPEC (solid squares) and long-life scientists in the
database (open squares). Solid lines represent fits to the data as
described in the text: log-normal distribution (gray line) with
$m=0.43\pm 0.01$ and $s=1.69\pm 0.01$, and distribution composed
of two power laws (black lines) one for small and intermediate
events ($\gamma=1.67\pm 0.01$) and the other for extreme events
($\gamma=2.87\pm 0.03$).}\label{fig3}
\end{figure*}

In Fig.~\ref{fig3} we have shown on logarithmic scales histograms
of the number of papers written by: all authors listed in INSPEC
and all long-life scientists in the database. As expected, both
distributions are highly skewed, and their fat-tails are due to
long-life scientists. One can also see that the distribution of
all authors regardless of their seniority is well described by the
log-normal distribution (\ref{lognormal}), which for reasons
elaborated by Sornette and Cont \cite{SornetteJPF97} (see also
\cite{FrischJPF97,LahEPJB98}) may be confused with distribution
having power law tail (\ref{powerlaw}). In the Fig.~\ref{fig3},
apart of the log-normal fit to our data, we have shown that
distribution composed of two power laws also fits our data very
well. Nevertheless, the exponents $\gamma$ for both regions of the
power law scaling significantly differ from the exponent
$\gamma\simeq 2$ predicted by Lotka.

The reported studies show that scientists differ enormously in the
number of papers they publish. Although, at present the fat-tailed
distributions are not so surprising for physicists as they were
$20$ years ago, the appearance of highly skewed distributions
characterizing scientific productivity is still strange since it
refers to scientific elite who undergone a rigorous selection
procedure and is expected to be more homogeneous. At the moment,
one may for example suggest that the noticed differences between
scientists may result from the heterogeneity of the analysed
sample (e.g. as is the case in nonextensivity driven by
fluctuations \cite{WilkPRL2000,BeckPRL2001}). To be ahead of these
suggestions, in the following we will concentrate on analysis of
$T$-cohorts, as they were characterized at the beginning of this
section. Although, the approach makes our data more homogeneous,
we are aware that it still does not take into account other
factors which influence scientific productivity (e.g. access to
resources which facilitate research or geopolitical conditions).
In the next section we will try to convince the readership that
the effect of those omitted factors may be understood in terms of
a single function having the same meaning as density of states in
equilibrium statistical physics.

\begin{figure*}
\centerline{\epsfig{file=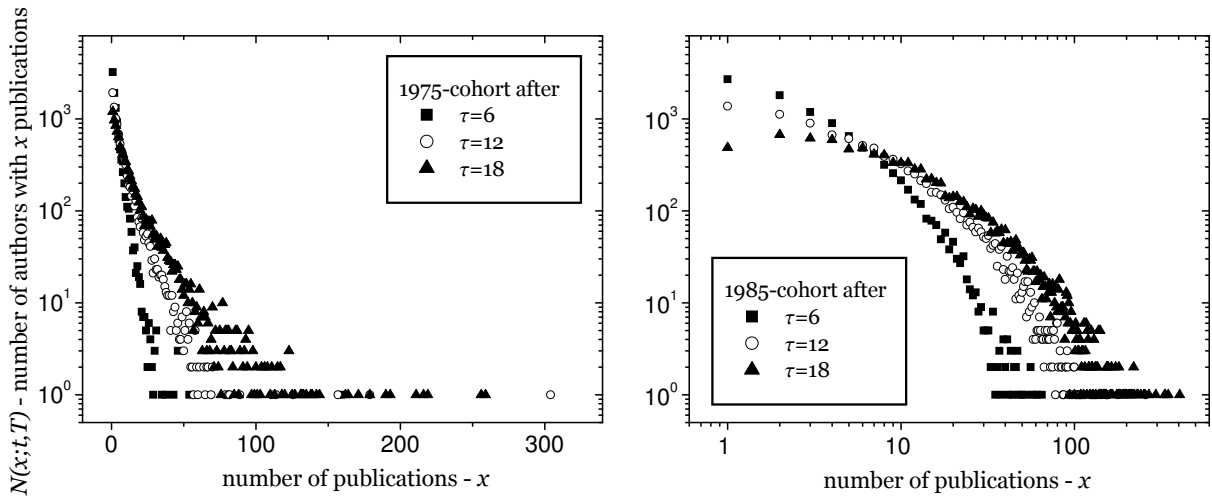,angle=0,width=16cm}}
\caption{Histograms of scientific productivity $N(x;t,T)$
characterizing cohorts of long-life scientists, who started to
publish in a given year $T=1975$ or $1985$, and
$\tau=t-T=6,12,18$. (Detailed description of the figure is given
in the text.)}\label{fig4}
\end{figure*}

Due to our approach, whatever differences are observed among
$T-$scientists they can be logically decomposed into only two
sorts: (i) life-course differences, which are the effects of
biological and social aging, and (ii) cohort differences, which
are differences between cohorts at comparable points in career
history. According to our knowledge the only similar analysis of
scientific productivity was performed by Allison and Stewart
\cite{Allison1974}, who analysed a sample of U.S. scientists in
university departments offering advanced degrees in biology,
chemistry, physics and mathematics. The authors divided the sample
into $8$ age strata by the number of years since Ph.D.,
representing different cohorts at different points during their
career history. Unfortunately, lacking longitudinal data the
authors were only able to observe life-course differences among
scientists, assuming that cohort differences are negligible.

\begin{table}[hp]
    \centering
        \begin{tabular}{cccccccc}
          \hline\hline
            $T$-cohort & $a$ & $b$ & $A$ & $B$ & $C$ & $E$ & $\tau_1$\\ \hline
            $\;1975\;\;$ & $\;0.025\;\;$ & $\;0.39\;\;$ & $\;0.06\;\;$ & $-\;1.02\;\;$ & $\;2.86\;\;$ & $\;0.48\;\;$ & $\;-7.24\;$ \\
            $1977$ & $0.028$ & $0.40$ & $0.03$ & $-1.47$ & $3.09$ & $0.86$ & $-7.49$ \\
            $1979$ & $0.035$ & $0.37$ & $0.06$ & $-0.97$ & $3.00$ & $0.58$ & $-5.38$ \\
            $1981$ & $0.048$ & $0.36$ & $0.01$ & $-2.15$ & $3.50$ & $3.20$ & $-4.60$ \\
            $1983$ & $0.055$ & $0.39$ & $0.01$ & $-2.38$ & $3.63$ & $3.37$ & $-4.53$ \\
            $1985$ & $0.066$ & $0.42$ & $0.04$ & $-1.38$ & $3.26$ & $1.31$ & $-3.64$ \\
            $1987$ & $0.119$ & $0.35$ & $0.07$ & $-1.36$ & $3.25$ & $1.36$ & $-1.80$ \\
            \hline\hline
        \end{tabular}
    \caption{Values of parameters $a,b,A,B,C,E,\tau_1$ for a few $T$-cohorts. See Eqs.~(\ref{Eqfig5a}),
    (\ref{Eqfig5b}), and (\ref{theta}).}
    \label{tab1}
\end{table}

In Fig.~\ref{fig4} we have presented how the histogram of
scientific productivity $N(x;t,T)$ depends on time $t$ as a
$T$-cohort ages. In general, the scenario is the same for all
analysed $T$-cohorts: $N(x;t,T)$ changes from almost exponential
(when a cohort contains young scientists) to clearly fat-tailed
(when the same cohort consists of mature researchers). The results
exemplify life-course differences among long-life scientists, and
in some sense confirm the so-called hypothesis of {\it
accumulative advantage} \cite{Allison1974}, which claims that due
to a variety social and other mechanisms productive scientists are
likely to be even more productive in the future, whereas those who
produce little original work are likely to decline further in
their productivity.

\begin{figure*}
\centerline{\epsfig{file=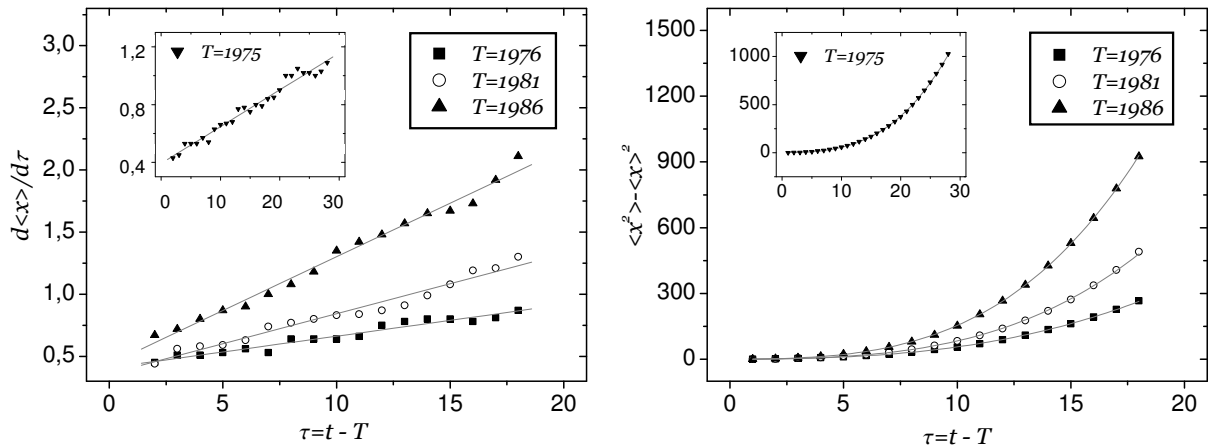,angle=0,width=16cm}}
\caption{Change of the average productivity $d\langle
x\rangle/d\tau$, and the variance $\langle x^2\rangle-\langle
x\rangle^2$ of cohorts' productivity distributions $N(x;t,T)$
versus seniority $\tau=t-T$. Points represent real data retrieved
from the INSPEC database, whereas solid lined express numerical
fits according to Eqs. (\ref{Eqfig5a}) and (\ref{Eqfig5b}).
(Detailed description of the figure is given in the
text.)}\label{fig5}
\end{figure*}

In order to examine cohort differences we have analysed how the
average $\langle x\rangle$ and the variance $\langle
x^2\rangle-\langle x\rangle^2$ of the distribution $N(x;t,T)$
depend on the cohort parameter $T=1975,\dots,1987$, and how they
change over time $t$. We have found that the parameters are
well-defined increasing functions of time (see Fig.~\ref{fig5})
\begin{equation}\label{Eqfig5a}
\frac{\partial\langle x\rangle}{\partial\tau}=a\tau+b,
\end{equation}
and
\begin{equation}\label{Eqfig5b}
\langle x^2\rangle-\langle x\rangle^2= A\left(\tau-B\right)^{C},
\end{equation}
where $\tau=t-T$ and $a,b,A,B,C$ depend on $T$ (see
Tab.~\ref{tab1}).

At the moment, it is worth to mention that although our analysis
encompasses only $18$ initial years of cohorts' history, we have
also verified the above relations for $28$ years of activity of
the oldest $1975$-cohort, finding excellent agreement with the
results obtained for other cohorts and for the shorter period of
time (see insets in Fig.~\ref{fig5}). Nevertheless, one should be
aware that even the most productive scientists in his/her
declining years slow down pace of working. According to Zhao
\cite{A2}, the optimal age for scientific productivity is between
$25$ and $45$, reaching the peak for researchers around $37$ (i.e.
about $18$ years since the beginning of the career). Similar
findings has been also reported by Kyvik \cite{A3}, who found that
publishing activity reaches a peak in the $45-49$-year-old age
group and declines by about $30\%$ among researchers over $60$
years old. Summing up, in the light of previous results on the
relation between age and productivity, findings reported in our
paper apply to scientists in the most prolific period of their
career.

\begin{figure} \epsfxsize=7.5cm \epsfbox{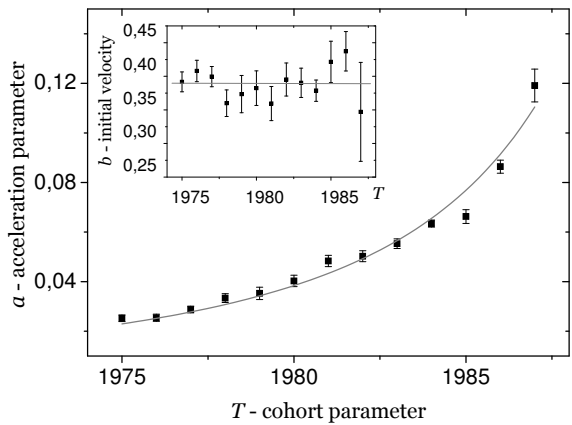}
\caption{Acceleration parameter $a$ and initial velocity $b$
versus cohort parameter $T$. As previously, points represent data
retrieved from INSPEC, whereas solid lines express trend in the
data.}\label{fig6}
\end{figure}

Now, let us briefly comment on the relations (\ref{Eqfig5a}) and
(\ref{Eqfig5b}). First, note that the linear dependence on
seniority $\tau$ in Eq.~(\ref{Eqfig5b}) implies that an average
representative of each cohort possesses an acceleration parameter
$a$, which is fixed during the whole scientific career. Moreover,
the parameter increases with $T$ (cf. Tab.~\ref{tab1} and
Fig.~\ref{fig6}), certifying that younger (in terms of $T$)
scientists are better skilled to produce more papers than their
older colleagues at the same point of the scientific career. It is
a matter of debate whether the differences in $a$ are due to
better adaptation of young people to technological achievements
(i.e. computers and the Internet), or they result from the rough
competition between researchers, and are one of syndromes of the
publish-or-perish phenomenon. In the next section, exploiting
relations (\ref{Eqfig5a}) and (\ref{Eqfig5b}), we will show that
regardless of the reasoning the explanation of accelerated
productivity naturally emerges as a result of treatment of the
scientific community by means of methods borrowed from equilibrium
statistical physics.

\section{Theoretical approach to scientific productivity -
density of states underlying scientific community}

In sociometrics, explanations of highly skewed histograms of
scientific productivity $N(x)$ (see Fig.~\ref{fig3}) are generally
of two (not necessarily exclusive) types \cite{Cole1973}. The {\it
sacred spark} (i.e. {\it heterogeneity}) hypothesis says that the
observed discrepancies in scientific productivity originate in
substantial, predominated differences among scientists in their
ability and motivation to do creative research, while the {\it
accumulative advantage} (i.e. {\it reinforcement}) hypothesis
\cite{Merton1968,Allison1974} claims that due to a variety of
social and other mechanisms productive scientists are likely to be
even more productive in the future. According to the first
hypothesis, skewed distributions of hidden attributes
characterizing scientists naturally lead to skewed distribution of
productivity, whereas the second hypothesis argues that the
observed fat-tailed histogram $N(x)$ results from sophisticated
stochastic processes underlying scientific productivity (see e.g.
\cite{Shockley1957,Simon1957}).

In this section we will present an alternative explanation of the
skewed productivity distributions. Since we have already noticed
that the fat-tail of the distribution $P(x)=N(x)/N$ characterizing
the set of all authors listed in INSPEC is due to long-life
scientists (c.f. Fig.~\ref{fig3}), in the following we shall only
concentrate on distributions $P(x;t,T)=N(x;t,T)/N(T)$
characterizing $T-$cohorts (see Fig.~\ref{fig4}). In order to
describe the scientific community, we will exploit the maximum
entropy principle \cite{Jaynes1957a,Jaynes1957b}, and we will
adopt some of the fundamental concepts from equilibrium
statistical mechanics (like statistical ensemble, phase space, and
density of states). We will also argue, that our approach does not
contradict the sociological hypothesis mentioned at the beginning
of the section.

In physics, the notion of statistical ensemble means a very large
number of mental copies of the same system taken all at once, each
of which representing a possible state that the real system might
be in. When the ensemble is properly chosen it should satisfy the
ergodicity condition, which guarantees that the average of a
thermodynamic quantity across the members of the ensemble is the
same as the time-average of the quantity for a single system.

In our approach we will identify a representative of a given
$T$-cohort with a physical system, and we will try to describe
such a system (i.e. a long-life scientist) in terms of statistical
physics. Since (at least now) we do not have access to parallel
worlds, in our approach a large group of copies of the same
scientist will be replaced with a large set of {\it
macroscopically} similar long-life scientists, i.e. scientists
belonging to the same $T$-cohort, and taken at a given point in
their scientific career $\tau=t-T$. Here, the assumption of
macroscopic similarity means that the considered scientists are
{\it exposed} to the same external field (influence)
$\theta(t,T)$, which forces (motivates) scientists to publish an
average number of publications $\langle x\rangle(t,T)$. The
external field (influence) $\theta$ has the same meaning as the
inverse temperature $\beta=(kT)^{-1}$ which determines the average
energy $\langle E\rangle$ in the canonical ensemble
\cite{JaynesWIKI}.

Now, suppose that one would like to establish probability
distribution $P(\Omega)$ over a given $T-$cohort at time $t$,
where
\begin{equation}\label{omega}
\Omega=\{y_1,y_2,\dots,y_n\}
\end{equation}
stands for states (i.e. microstates) of a single scientist, who
belongs to the considered cohort /ensemble. (Let us explain that
the parameters $y_i$ are coordinates of a hidden phase space
underlying the scientific community, and determining scientific
productivity
\begin{equation}\label{x}
x=x(\Omega)=x(y_1,y_2,\dots,y_n).
\end{equation}
Of course, there exists a number of such parameters, including:
research field, IQ level, age, number of coworkers, motivation,
funds etc., but as it turns out in the rest of this section a few
important findings about our ensembles may be obtained even
without detailed knowledge on the parameters.) Due to the maximum
entropy school of statistical physics initiated by Edwin T. Jaynes
in 1957 \cite{Jaynes1957a,Jaynes1957b}, the best choice for the
distribution $P(\Omega)$ is the one that maximizes the Shannon
entropy
\begin{equation}\label{entropy}
S=-\sum_\Omega P(\Omega)\ln P(\Omega),
\end{equation}
subject to the constraint
\begin{equation}\label{srx}
\langle x\rangle(t,T)=\sum_\Omega P(\Omega)x(\Omega),
\end{equation}
plus the normalization condition
\begin{equation}\label{norm}
\sum_\Omega P(\Omega)=1.
\end{equation}

The Lagrangian for the above problem is given by the below
expression
\begin{eqnarray}\nonumber
\mathcal{L}=&-&\sum_\Omega P(\Omega)\ln
P(\Omega)+\alpha(t,T)(1-\sum_\Omega
P(\Omega))\\&+&\theta(t,T)\left(\langle x\rangle(t,T)-\sum_\Omega
x(\Omega)P(\Omega)\right)\label{Lag},
\end{eqnarray}
where the multipliers $\theta(t,T)$ (external field) and
$\alpha(t,T)$ are to be determined by (\ref{srx}) and
(\ref{norm}). Differentiating $\mathcal{L}$ with respect to
$P(\Omega)$, and then equating the result to zero one gets the
desired probability distribution over the $T-$cohort
\begin{equation}\label{PO}
P(\Omega)=\frac{e^{-\theta(t,T)x(\Omega)}}{Z(t,T)},
\end{equation}
where $Z(t,T)$ represents the partition function (normalization
constant), and
\begin{equation}\label{Z}
Z(t,T)=\sum_\Omega e^{-\theta(t,T)x(\Omega)}=e^{\alpha(t,T)+1}.
\end{equation}

\begin{figure} \epsfxsize=7.5cm \epsfbox{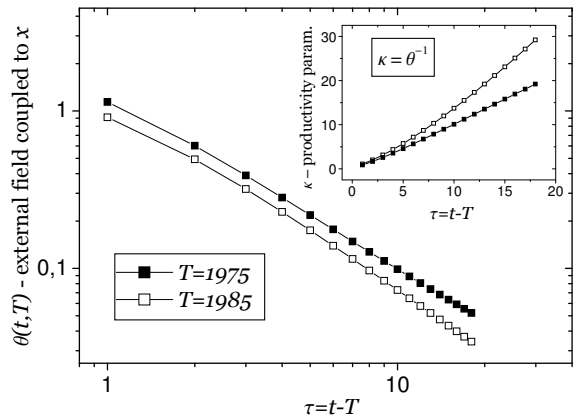}
\caption{Main stage: external field (influence) $\theta(t,T)$
versus seniority $\tau=t-T$ for two cohorts $T=1975$ and $T=1985$.
Subset: productivity parameter defined as $\kappa=\theta^{-1}$
versus $\tau$ for the same cohorts.}\label{fig7}
\end{figure}

Before we proceed further, let us make two comments here. First,
since each $T-$cohort changes over time $t$ a sceptic may bring
the validity of our {\it equilibrium} approach into question. In
order to justify the approach we assume that time dependence of
$T$-cohorts may be considered in terms of {\it quasistatic}
equilibrium process. (Let us remind that in a quasistatic process,
due to sufficiently slow dynamics, a system is considered to cross
from one equilibrium state to another.) The assumption allow us to
treat each $T-$cohort in separate years $t>T$ as an equilibrium
system. The second comment relates to {\it ergodicity} of our
ensembles. In statistical physics the ergodic hypothesis says
that, over long periods of time, the time spent in some region of
the phase space corresponding to microstates with the same energy
is proportional to the volume of this region, i.e. that all
accessible microstates $\Omega$ are equally probable over long
period of time. Equivalently, the hypothesis says that time
average and average over the statistical ensemble are the same. In
the case of long-life scientists, we may only speculate about the
underlying phase space, its dimensionality and coordinates
(\ref{omega}). Even if we were able to enumerate most of
significant coordinates characterizing such scientists, surely a
part of these coordinates, including e.g. motivation, would be
impossible to quantify. Summarizing, given the above and other
difficulties it appears impossible to verify the ergodic
hypothesis for our ensembles, and the question - if ergodicity is
fulfilled here - remains open.

\begin{figure} \epsfxsize=7.5cm \epsfbox{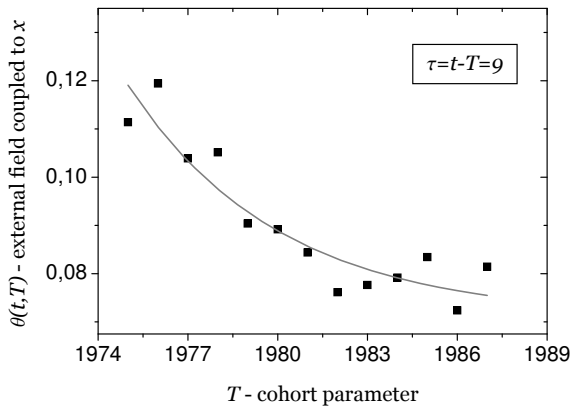}
\caption{Differences between cohorts. External field $\theta(t,T)$
coupled to the number of publications $x$ versus the cohort
parameter $T$ for $\tau=t-T=9$. The solid line stands for trend in
the empirical data.}\label{fig7a}
\end{figure}

\begin{figure*}
\centerline{\epsfig{file=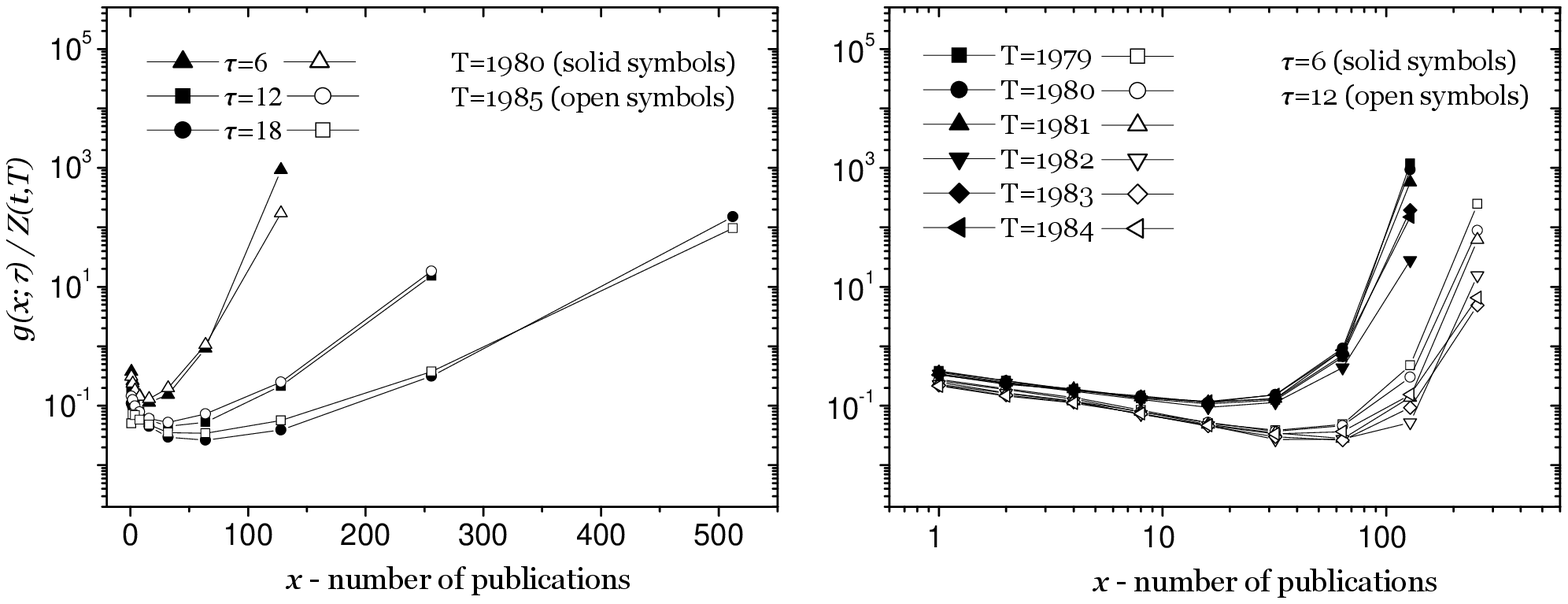,angle=0,width=16cm}}
\caption{Density of states functions $g(x;\tau)$ underlying
different $T-$cohorts at different stages of their scientific
career $\tau$.}\label{fig8}
\end{figure*}

Now, having the theoretical framework we are in a position to
analyze how the external field $\theta(t,T)$ influencing
scientists depends on $T$, and how it changes over time $t$. In
order to calculate the parameter we use the
fluctuation-dissipation relation
\begin{equation}\label{FDT}
\langle x^2\rangle-\langle x\rangle^2=-\frac{\partial\langle x
\rangle}{\partial\theta}=-\frac{\partial\langle x\rangle}{\partial
\tau}\left(\frac{\partial\theta}{\partial\tau}\right)^{-1},
\end{equation}
which may be simply derived from $P(\Omega)$ (\ref{PO}). (Keep in
mind that the ensemble averages $\langle x\rangle$ and $\langle
x^2\rangle$, and also $\theta$ depend on both $t$ and $T$.) At the
moment, note that in the previous section we have already found
empirical relations corresponding to both sides of the last
formula. Inserting the relations (\ref{Eqfig5a}) and
(\ref{Eqfig5b}) into (\ref{FDT}), after some algebra one obtains
\begin{eqnarray}\label{theta}
\theta(t,T)&=&-\int_{\tau_0}^\tau\frac{a\xi+b}{A(\xi-B)^C}d\xi\\\nonumber
&=&E(\tau-B)^{1-C}(\tau-\tau_1)+D,
\end{eqnarray}
where parameters $a,b,A,B,C,D$ depend on $T$, whereas $E,\tau_1$
are functions of these parameters (see Tab.~\ref{tab1}).

In Fig.~\ref{fig7} we have presented how the external field
$\theta(t,T)$ changes over seniority $\tau$. Since the field
conjugates to the cumulative number of publications, its
decreasing character indicates that small values of the field
correspond to large productivity, and vice versa - large fields
induce small productivity. (The inverse of $\theta$, i.e.
$\kappa=\theta^{-1}$, stands for a productivity field which has
more obvious sociological interpretation: larger $\kappa$ enforces
larger number of papers. See inset in Fig.~\ref{fig7}.) Having in
mind the reverse relationship between $\theta$ and the number of
publications $x$, one can argue that the constant of integration
$D$ in (\ref{theta}) must be equal to zero. The reasoning behind
the statement is the following. Given that the considered
long-life scientists never die, still being in the most prolific
period of their career, one may simply imagine that in the limit
of $\tau\simeq t\rightarrow\infty$ the total number of
publications produced by these scientists must approach infinity,
what corresponds to $\theta(\infty,T)=0$, and respectively
$D(T)=0$.

The above results allow us to further investigate differences
between $T$-cohorts. Comparing values of the external field
$\theta(t,T)$ influencing $T$-scientists at the same point
$\tau=t-T$ in their scientific career, one can show that the field
is a decreasing function of $T$ (see Fig.~\ref{fig7a}). (We have
also checked that the decreasing character of $\theta(T+\tau,T)$
versus $T$ holds for every value of $\tau=1,2,\dots,18$.) The
above stems from the fact that younger (in terms of $T$)
scientists publish more than their older colleagues at the same
age. The interesting point here is that statistical physics allows
to describe the phenomenon in terms of changing external field,
which leads to accelerated productivity as described in the
previous section.

In order to finalize our theoretical approach to scientific
productivity we should explain the mutual relationship between the
theoretical distribution $P(\Omega)$ (\ref{PO}) and the empirical
distribution $P(x;t,T)$ (see Fig.~\ref{fig4}). Thus, since the two
distributions apply to the same ensembles there should exist a
possibility to cross from one distribution to the other. Such a
possibility appears due to the density of states function
$g(x;t,T)$, which expresses the number of allowed states $\Omega$
(cf. Eq.~\ref{omega}) that scientists may be in, given that the
number of publications corresponding to these states equals $x$
(\ref{x}). Using the concept of the density of states one can
write
\begin{equation}\label{g1}
P(x(\Omega);t,T)=g(x;t,T)P(\Omega),
\end{equation}
and respectively the empirical function $g(x;t,T)$, correct to the
multiplicative factor $Z(t,T)$, may be obtained from the below
expression
\begin{equation}\label{g3}
\frac{g(x;t,T)}{Z(t,T)}=P(x;t,T)e^{\theta(t,T)x}.
\end{equation}

In Fig.~\ref{fig8} we have presented how the empirical density of
states $g(x;t,T)$ depends on $x$. The most striking feature about
$g(x;t,T)$ is that it does not depend separately on time $t$ and
$T$, but it depends on their difference $\tau=t-T$ (cf. bunches of
curves shown in the figure)
\begin{equation}\label{g4}
g(x;t,T)\equiv g(x;\tau).
\end{equation}
The above means that the density of states is an inherent
characteristic describing researchers of a given seniority $\tau$.
It also certifies that the parameter $\theta(t,T)$ (\ref{theta})
has the meaning of an external field, which is only responsible
for filling of corresponding states (\ref{omega}) in the hidden
phase space underlying scientific community. The analogy between
our parameter $\theta$ and the inverse temperature $\beta$ in the
canonical ensemble is indeed very close. External conditions
expressed by the field $\theta$ do not change the considered
system, which in our case corresponds to a scientist characterized
by a given value of $\tau$. They only influence the probability
(\ref{PO}) of realization of a state corresponding to a given
productivity $x$ (\ref{x}). In particular, the findings allow us
to say that representatives of younger cohorts usually coauthor
much more articles than their counterparts (in terms of the same
$\tau$) belonging to older cohorts. It means that due to external
requirements (which we interpret as publish-or-perish phenomenon)
representatives of younger cohorts are skilled (forced) to
contribute more articles.

\begin{figure} \epsfxsize=8.5cm \epsfbox{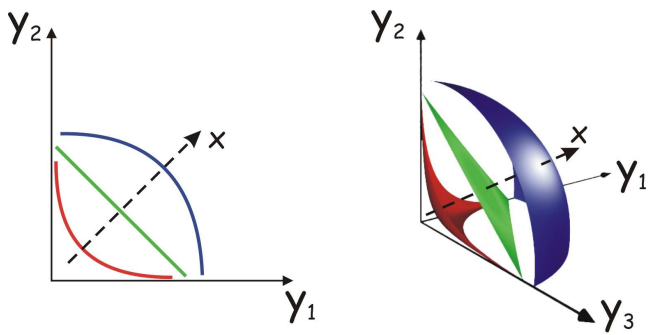}
\caption{Examples of phase trajectories $x(\Omega)$ in the space
of scientific motivators $\Omega=\{y_1,y_2,\dots,y_n\}$ resulting
in the corresponding shape of $g(x;\tau)$. (Detailed description
of the figure is given in the text.)}\label{fig9}
\end{figure}

Finally, before we proceed to conclusions let us briefly comment
on the shape of the function $g(x;\tau)$ (see Fig.~\ref{fig8}).
The function monotonically decreases for small and quickly
increases for large values of $x$, having the characteristic
minimum for intermediate $x$. One can argue that the corresponding
curvature of $g(x;\tau)$ may result from topological requirements
imposed by the relation $x(\Omega)$ (\ref{x}) on the hidden space
$\Omega=\{y_1,y_2,\dots,y_n\}$ (\ref{omega}). A simple but still
reasonable example of such a relation is graphically presented in
Fig.~\ref{fig9}. (Although the figure presents only two- and
three-dimensional phase spaces the below reasoning also holds for
higher dimensions.) In the figure, the direction of the dashed
lines expresses growing number of publications $x$, whereas the
area of the $n-$dimensional hypersurface is proportional to the
number of states $g(x;\tau)$ of a given value of $x$. As one can
see, the hypersurfaces $x(\Omega)$ corresponding to increasing
values of $x$ change from convex to concave. The feature leads to
the minimum in the density of states function, and has a nice
sociological interpretation.

In order to outline the mentioned sociological interpretation, let
us assume that all motivators $y_i$ influencing scientific
productivity have some minimal values. Such an assumption seems to
bee natural since one can not get salary lower than a certain
limit, and it is impossible to possess negative number of
coworkers. On the other hand, there are no upper limits for these
parameters. We are not even in a position to guess their units. It
follows that for visualization purposes all motivators may be
limited to their positive values, as shown in Fig.~\ref{fig9}.
Now, in order to justify the suggested convex character of the
hypersurface $x(\Omega)$ representing small values of $x$, one can
argue that it corresponds to the leading role of one selected
motivator $y_i$, and insignificant role of other parameters
$y_{j\neq i}$. In some sense, such a naive thinking on factors
influencing scientists is consistent with a common experience
stating that in early stages of career the only one factor makes
motivation for scientific activity (e.g. satisfaction). Along with
growing $x$ other motivators start to play a role (e.g.
recognition and being in power), what may be expressed by the
mentioned {\it convex-to-concave} crossover.

\section{Summary}

In this paper we have attempted to provide a quantitative approach
to the publish-or-perish phenomenon, which refers to the pressure
to constantly publish work in order to further or sustain one's
scientific career. Using data retrieved from the INSPEC database
we have quantitatively discussed a few syndromes of the
phenomenon, including continuous growth of rate of scientific
productivity, and continuously decreasing percentage of those
scientists who stay in science for a long time. Methods of
equilibrium statistical physics have been applied for the
analysis. We have shown that the observed fat-tailed distributions
of the total number of papers $x$ authored by scientists may
result from a specific shape of the density of states function
$g(x;\tau)$ underlying scientific community. We have also argued
that although different generations of scientists are
characterized by different productivity patterns, the function
$g(x;\tau)$ is inherent to researchers of a given seniority
$\tau$, and the publish-or-perish phenomenon may be quantitatively
characterized by the only one time- and generation- dependent
parameter $\theta$, which has the meaning of an external field
influencing researchers.

\section{Acknowledgments}

We thank Andrea Scharnhorst from Virtual Knowledge Studio for
Humanities and Social Sciences at Royal Netherlands Academy of
Arts and Sciences, and Loet Leydesdorf from Department of Science
and Technology Dynamics at University of Amsterdam for useful
comments and suggestions.

The work was funded in part by the European Commission Project
CREEN FP6-2003-NEST-Path-012864 (P.F.), and by the Ministry of
Education and Science in Poland under Grant 134/E-365/6, PR UE/DIE
239/2005-2007 (A.F. and J.A.H.). A.F. also acknowledges financial
support from the Foundation for Polish Science (FNP 2006).

\end{document}